\documentstyle[prl,twocolumn,aps]{revtex}

\input epsf
\begin{document}
\draft

\twocolumn[\hsize\textwidth\columnwidth\hsize\csname@twocolumnfalse\endcsname
\title{Noise enhanced stability of periodically
driven metastable states} 

\author{N. V.
Agudov$^{1,2}$ and B. Spagnolo$^{1}$}

\address{$^{1}$INFM, Unit\'{a} di
Palermo, and Dipartimento  di Fisica e
Tecnologie Relative, Universit\'{a} di
Palermo,\\ Viale delle Scienze, I-90128
Palermo, Italia\\ $^{2}$ Radiophysical
Department, State University  of Nizhny
Novgorod, $23$ Gagarin Ave.,\\  Nizhny
Novgorod $603600$, Russia} 

\date{\today} 
\maketitle

\begin{abstract}
We study the effect of noise enhanced stability of
periodically driven metastable states in a system
described by piecewise linear potential. 
We find that the growing
of the average escape time with the intensity of the noise is
depending on the initial condition of the system.
 We analytically obtain the condition for
the noise enhanced stability effect and verify it by numerical simulations.
\end{abstract}


\vskip1pc]
Escape from a metastable state is a phenomenon observed 
in several
scientific areas. Among them there are the theory of diffusion 
in solids,
chemical kinetics and
transport in
complex systems \cite{Hanggi90}.

The mean first passage time (MFPT) of a Brownian particle moving 
in potential fields usually decreases with noise intensity
according to the Kramers formula \cite{Kramers40} or some universal 
scaling function of the system parameters \cite{Ferdi91,Maxi97}.
The dependence on the noise intensity of the MFPT for
metastable and unstable systems was revealed to have 
resonance character first noted by Hirsch et al. 
\cite{Hir82} and then observed in
different physical systems 
\cite{WeisCas,Nik95-99_And96,RB96-00,%
LucaF97,Mahato97-99,Wac98-99_Mielke00}.
The most important conclusion of these studies is that
the noise can modify the stability of the system in a
counter-intuitive way. The system remains in the
metastable state for a longer time than in the deterministic 
case and the escape time has a maximum at some noise
intensity.
Noise enhanced
stability (NES) originally found numerically by Dayan et al.
\cite{WeisCas},
was observed experimentally in a tunnel diode
by Mantegna and Spagnolo\cite{RB96-00}.
More recently it was found that the
noise induced slowing down \cite{Mahato97-99}
and the noise-induced stabilization  \cite{Wac98-99_Mielke00}
are related to NES phenomenon \cite{RB96-00}.

Some questions arise from previous studies:
(i) What is the reason of the increase of the
average escape time with the noise intensity?
(ii) What about the condition for which the NES effect
takes place?
To answer both questions we investigate the escape time
from a periodically driven metastable state 
for a piecewise linear potential.
We study the nature of this
phenomenon analytically. We find that for fixed potential 
the decay time of unstable initial state can be dramatically
increased by the presence of a small noise 
depending on the initial condition of
the system. We obtain the condition for the NES effect, as an explicit
relation between the driving frequency and the parameters of the
potential.

We consider the model of overdamped Brownian motion
described by the equation

\begin{equation}\label{lan}
{dx\over dt}=-{\partial U(x)\over \partial x} + F(t) + \xi(t),
\end{equation}
where  $\xi(t)$
is the white Gaussian noise with zero mean,
$\langle\xi(t)\xi(t+\tau)\rangle=2q\delta(\tau)$, 
$F(t)$ is the dichotomous driving
force and $U(x)$ is a potential profile defined as 

\begin{equation}\label{U}
U(x)=\left\{
\begin{array}{ll}
\infty,&x=0\\
hx,&0<x\leq\ell\\
E-k(x-\ell), & \ell\leq x<b
\end{array},
\right.
\end{equation}
with $E=h\ell$. Specifically we assume $k>0$
and $|F(t)|<k$.

First we consider the system governed by Eq.~(\ref{lan}) with
$F(t)=0$ and potential $(2)$ with arbitrary $h$. 
If $h>0$ ($E>0$)
the states at $x<\ell$ become metastable. 
The
exact expression of the MFPT
from initial position $x_0$ to boundary $b$
is known
for the case $F(t)=0$

\begin{equation}\label{T} 
\tau(x_0,b,q)={1\over q}\int_{x_0}^b
e^{u(z)}\int_{-\infty}^z e^{-u(y)}dy dz, 
\end{equation}
where $u(x)=U(x)/q$ is a dimentionless potential profile. 
Evidently in physical systems we can not observe the
microscopic initial conditions. However the MFPT of
Eq.(\ref{T}) it is sufficient to obtain the MFPT with
arbitrary macroscopic initial distribution by simple
integration.
Therefore further we study $\tau(x_0,b,q)$ because it contains
the full information about the system.
If
$0<x_0<\ell$, the decay time $\tau=\tau_1$ for the potential
profile of Eq. (\ref{U}) is 

\begin{eqnarray}\label{T1}
\tau_1(x_0,b,q)={b-\ell\over k} -{\ell-x_0\over h}+
\frac{q(h + k)}{h^2 k} e^{E/q}\\
-{q\over h^2}e^{hx_0/q}-
 \frac{h + k}{k^2 h} \left(
1-e^{-A/q}\right)-{q\over kh}e^{(E-A)/q}.
\nonumber
\end{eqnarray}
If $\ell<x_0<b$, the decay time $\tau=\tau_2$ is

\begin{eqnarray}\label{T2q}
\tau_2(x_0,b,q)={1\over k}\left[
\frac{q(h + k)}{hk}
\left(e^{-A/q}-e^{-\Delta E/q}\right)\right.\\
\left. + b-x_0+{q\over h}
\left(e^{(E-\Delta E)/q}-
e^{(E-A)/q}\right)\right].
\nonumber
\end{eqnarray}
Here $A=k(b-\ell)$, and  $\Delta E=k(x_0-\ell)$. 

These expressions  show that
at large noise intensity 
the decay time $\tau(q)$ decreases with noise 
as $1/q$ for arbitrary $h$. 
When the noise intensity is small, $q \ll |E|$ the 
influence of the potential
barrier becomes significant. 
For $E<0$ and $h<0$, the barrier is absent,
and the NES effect, also known as noise delayed decay, 
appears when $|h|<k$ and
$x_0$ is near $\ell$ \cite{Nik95-99_And96}.
Indeed, for $q\ll |E|$,$A$, $|E-A|$ 
the decay time $\tau_2(\ell,b,q)$
grows with noise temperature
$q$. When $x_0 = 0$, the decay time always decreases
with $q$. 

In the case with potential barrier ($E>0$, $h>0$) the 
escape time depends on the
initial position of the particle with respect to the 
potential barrier. 
When the  particle is within the potential well ($x_0<\ell$)
 the decay time of 
metastable state increases infinitely when $q\to0$, because 
if the noise is
absent, the particle can never surmount the potential
 barrier. For $q\ll E$
 the decay time (\ref{T1}) coincides with the Kramers' time,
 which in this case
reads 

\begin{equation}\label{T1k}
\tau_1(x_0,b,q)\simeq \tau_k=
\frac{q(h + k)}{h^2 k} e^{E/q}
\mathop\rightarrow_{q\to0}\infty.
\end{equation}
When $\ell<x_0<b$ the initial state of the particle 
is unstable. 
In the absence of noise, the
escape time from this unstable state  is a finite value: 
$\tau_2(x_0,b,0)=(b-x_0)/k$, which does not depend on
the potential well. When we add the noise the influence 
of potential well
becomes important. It follows from the exact expression 
of Eq.(\ref{T2q}) that the MFPT rises
to infinity if $\Delta E<E$

\begin{equation}\label{T2nes} 
\tau_2(x_0,b,q)\simeq {q\over kh} 
e^{(E-\Delta E)/q}
\mathop\rightarrow_{q\to0}\infty,  
\end{equation}
while for the case $\xi(t)=0$ we have the decay time
obtained from the deterministic Eq.~(\ref{lan}), i.~e. the MFPT
 has a
singularity at $q=0$, when $\Delta E<E$. 
From a physical 
point of view this
singularity can be explained as follows:
 When the particle is initially
located in the region $\ell<x_0<b$, a small quantity of noise added 
in the system can
eventually push the particle into potential well. Then, 
the particle will be
trapped there for a long time because the well is very deep. 
This type
of trajectories of the Brownian particles therefore leads to  
a big "tail" in the First
Passage Time Distribution (FPTD) $w(t)$. 
If the potential well
is very deep, namely, $E>\Delta E$ the trapping time is so 
long that the integral for
MFPT

\begin{equation}\label{fpt}
\tau=\int_0^\infty t w(t) dt
\end{equation}
diverges when $q\to0$. The FPTD obeys the 
backward Fokker-Planck equation
and can be obtained for piece-wise linear
potential using the Laplace
transform method \cite{Nik93}. 
The Laplace transform of the FPTD for our 
potential, when $h=k$ and $\ell<x_0<b$,
reads

\begin{equation}\label{FPTL}
\hat w(s)=\int_0^\infty w(t) e^{-st} dt=
{B(x_0,s)\over B(b,s)},
\end{equation}
where
$$
B(x,s)=c\mu e^{\lambda x-2\mu\ell}-c\lambda e^{-\mu x+2\lambda\ell}
+\gamma^2 e^{-\beta} \left(
e^{-\mu x}+e^{\lambda x}
\right)
$$
and $c=k/2q$, $\gamma=\sqrt{s/q}$, $p=\sqrt{c^2+\gamma^2}$, $\lambda=p-c$,
$\mu=p+c$, $\beta=k\ell/q$. Using
the limit theorems of Laplace transform, we can obtain 
from Eq.(\ref{FPTL}) the
asymptotic expression for $t\to\infty$ and $q\ll E$

\begin{eqnarray}\nonumber
&w_{as}(t)= G(t)
\left[
1+{1\over2}
\left(
{t_1+{1\over2}\tau_k\over t_0}-
{t_1^2\over t_0 t}
\right)
e^{\left({-{4\theta(x_0)\theta(b)\over\tau_k t}}\right)}
\right],&\\ 
&G(t)={t_0\over\sqrt{\pi\tau_k t^3}}
\exp\left(
-{(t-t_0)^2\over\tau_k t}
\right),&
\nonumber
\end{eqnarray}
where $\theta(x)=(x-2\ell)/k$, $t_0=\theta(b)-\theta(x_0)$, 
$t_1=\theta(b)+\theta(x_0)$, and $\tau_k=4q/k^2$. 
The time dependence of $w_{as}(t)$ 
is shown in Fig.~1 for three
different values of the initial position
$x_0=2.5$ ($\Delta E>E$), $x_0=2$
($\Delta E=E$),  and $x_0=1.2$
($\Delta E<E$). One can see that the 
tail of the FPTD
rises when $\Delta E$ decreases.
If $\Delta E>E$, the 
trapping time in the well is not very long and the integral 
of Eq.(\ref{fpt}) 
always converges.
Nevertheless, the average decay time increases with small 
noise, reach the
maximum and, then, decreases.
The plots of $\tau_2(q)$ for
various relations between $\Delta E$ and $E$ are 
shown in Fig.~2.
The main conclusion from the above analysis is that 
the strong effect of
NES can appear for the fixed potential profile with barrier, if 
the initial probability distribution is located
within the interval ($l,b$), i. e. in an unstable state
beyond the potential well. The physical system can be
brought in this nonequilibrium state by a sudden change of
control parameters. Examples of such situations include
spinodal decomposition in the dynamics of phase transitions
and the process of laser switch-on 
\cite{Maxi97,Fer84_Tito89}.
Such relaxation processes in the systems which 
are far from equilibrium attract now a great deal of
attention 
\cite{Mahato97-99,Wac98-99_Mielke00,Dyk99-Ha00_Ma01}. 

The main aim of this paper however is to
study the escape from the metastable state with an initial
distribution located within the potential well
in the presence of periodical driving. Further we
will apply the above results, obtained in the static
case, for analysis of periodical force effect.
Let's consider
the same potential profile $U(x)$ of Eq.(\ref{U}) but with $h=0$.
The driving force is $F(t) = a \nu (t)$, where $\nu (t)$ is 
the dichotomous signal switching between $\pm 1$ with period T
and $a$ is the amplitude.
We choose $\nu (t) = - 1$ (i.~e. the barrier is absent) for the first half of
period.
The exact expression for the MFPT
when potential varies  with time is unknown. 
Three recent papers \cite{Dyk99-Ha00_Ma01} develop a theory of
escape rates for periodically driven systems.
Smelyanskiy et al. consider the case of high potential barrier 
and small driving amplitude. Lehmann et al.
consider the regime of strong and moderately fast driving.
Maier and Stein analyse the crossover regime.
In all these cases the deterministic
escape time is infinity.  In the present work we consider 
the different regime of
strong  and moderately slow modulation
when the deterministic escape time is finite.

Therefore we first start our 
investigation from the deterministic case $q=0$ and secondly 
we define the 
condition for NES effect
when noise intensity is small.  
If $x_0=0$, the
Eq.~(\ref{lan}) has a periodic solution in
the deterministic regime 
for $T<2\ell/a$. 
In this case $x(t)<\ell$ for any $t$ and the particle
always  remains in the
metastable state. 
If the period is 

\begin{equation}\label{w}
T>2\ell/a
\end{equation}
the particle surmounts the potential barrier at time 
$t = \ell/a$. 
To obtain NES effect we should consider only the cases when the 
states are unstable
without noise \cite{RB96-00}. Consequently the first condition for 
the NES is 
given by Eq.(\ref{w}).  
It follows from the above analysis that the decay time 
in the presence of noise
strongly depends on potential barrier, namely, the barrier is
responsible for the strong increasing of decay time with noise. 
The exact
expressions for fixed potential show that increasing of decay 
time in the case
without barrier is much smaller 
and it appears only if the particle 
is near
the point $x_0=\ell$ (see Eq.s~(\ref{T1}) and (\ref{T2q})). 
Therefore 
it is important to consider two cases: (i) $\tau(0,b,0)<T/2$  and
(ii) $\tau(0,b,0)>T/2$. In the first case the modulation frequency 
is so low that
during the entire process of decay the potential profile $\Phi(x,t)$ 
has no
barrier.  In this case the average escape time
is not increased by the noise (see Eq.~(\ref{T1}) at $x_0 = 0$
and $h = -a$). In case of $\tau(0,b,0)>T/2$,  
average escape time can
be represented as follows

\begin{equation}\label{T+T}
\tau(0,b,0)={T\over2}+\tau(x(T/2),b,0)
\end{equation}
where $x(T/2)$ is the position of the particle at time $t=T/2$.
The conditions (\ref{w}) and $\tau(0,b,0)>T/2$ 
together mean that 

\begin{equation}\label{x1}
\ell<x(T/2)<b.
\end{equation}
Now we add a small quantity of noise into the system. 
The MFPT $\tau(0,b,q)$ can be written as 

\begin{equation}\label{sumMFPT}
\tau(0,b,q)=\tau(0,x_i,q)+\tau(x_i,b,q),
\end{equation}
where $x_i$ is an arbitrary point between $0$ and $b$.
For $x_i=x(T/2)$ and very small noise, the first term in the right
hand side of Eq.~(\ref{sumMFPT}) is approximately equal to the 
deterministic time: $\tau(0,x(T/2),q)\simeq\tau(0,x(T/2),0)=T/2$,
because the MFPT varies smoothly
with noise, when the barrier is absent (See Eq.~(\ref{T1})). In this
case: $\tau(0,b,q)\simeq T/2 + \tau(x(T/2),b,q)$, and
$\tau(x(T/2),b,q)\gg \tau(x(T/2),b,0)$ because of the potential
barrier, which makes the average escape time very large just for
$q\to0$ (see Eq.(\ref{T2nes})). As a result the decay time
$\tau(0,b,q)$
 will increase
with $q$ and the NES appears. (The decay time $\tau(0,b,q)$ will
not grow infinitely at $q\to0$ because the barrier exists only a
half of period.) Thus, we may conclude that the inequality
(\ref{x1}) must be the condition for NES. This inequality can be
rewritten as follows:

\begin{equation}\label{c1}
2{\ell\over a}<T<
2{ab+k\ell\over a(a+k)}.
\end{equation}

The inequality (\ref{c1}) and the condition $a<k$ give the area on
the $(T,a)$ plane where the NES effect takes place. In Fig.~3
we show this area for $k=1$, $\ell=2$, and $b=7$, and the results
of numerical simulations (shaded area).
We perform $3000$ different realizations of the decay process
$x(t)$ to
determine the average escape time for each couple of 
values of the amplitude $a$ and the 
period $T$ of the driving force. We consider more
than $100$ points on the $(T,a)$ plane. 
We find that within the area defined by inequality (\ref{c1}) 
the NES effect is
very strong: the average escape time increases more
than $10\%$ above the deterministic escape time. 
Outside this area and below the lower boundary the
deterministic decay time becomes infinite and the
NES disappears. In the presence of noise we obtain
Kramers-like behaviour.
This case was studied in detail by
Lehmann et al.\cite{Dyk99-Ha00_Ma01}. 
Inside the area the magnitude of the NES effect 
decreases from the lower to the
upper boundary.
Above the upper boundary the NES effect decreases
sharply. 
When the period $T$ and the amplitude $a$ 
of the driving force are chosen near
 the upper boundary of (\ref{c1}),
the potential barrier is very small or 
absent during the process of decay.
It explain why the effect is
very small, when we are near this boundary.
We also carried out simulations for $a>k$ 
and found the NES effect.
This parameter region however
gives less information about the 
mechanisms of the
NES effect from viewpoint of the interplay between the regular, 
random, and
periodical forces.
In this region in fact
the deterministic motion of the particle is 
characterized by oscillations and
the driving force
prevails the regular one described by 
the potential $U(x)$.

Therefore we conclude that the 
numerical simulations are in good agreement with 
the theory. The main 
mechanism of
the NES is defined correctly: it is the role of the
potential barrier
which 
appears after the
particle has crossed the point $\ell$ of maximal potential.
Consequently the inequality (\ref{x1}) is the
most general condition for NES effect because it can be applied
to a system described by an arbitrary potential with metastable state
and where $\ell$ is the x-coordinate of the maximum of the barrier 
appearing at $t = T/2$.   
The mechanism of NES explains why 
the FPT distributions
obtained in simulations and in experiments are multipeaked, 
periodic and with an exponential time decaying envelop 
\cite{WeisCas,RB96-00}.
The peaks appear only for small noise intensity,
where the NES effect occurs.
The first peak corresponds to the deterministic escape time.
The second peak arises because the small noise
provides the above considered inverse probability current which
moves some particles into the potential well.
The returned particles can escape only in one period.
Therefore the second peak is one period apart from the first one.
After each period we have the
same physical situation and as a consequence fewer particles
go back into the potential well.
Therefore the probability peaks have period $T$
and they decrease with time.
 The probabilities of escape are
independent and equal for successive oscillations of the potential.
So if the escape probability per oscillation is $p$,
the probability to escape at the {\em n}-th cycle is
$(1-p)^{n-1}p \simeq p e^{\alpha} e^{-\frac{\alpha t}{T}}$
where $-\alpha = \ln(1-p)$ and $n \simeq t/T$. Therefore
the magnitude of the FPT peaks are 
exponentially decreasing with time.

We acknowledge Dr. R. N. Mantegna 
for carefully reading the manuscript. This work was supported 
by INFM, MURST and by RFBR (Proj. Nos. 99-02-17544, 00-15-96620).

\small
\noindent
FIG. 1. Asymptotic behaviour of the FPTD $w_{as}(t)$ 
versus time $t$ for three values of the initial
position: $x_o = 2.5$ ($\Delta E>E$ dot-dashed line), 
$x_0=2$ ($\Delta E=E$ full line), 
and $x_0=1.2$ ($\Delta E<E$ dashed line). 
The parameters are: $b = 3, k = h = 1, l =1, 
q = 0.01$.

\noindent
FIG. 2. The normalized decay time $\tau_2(q)/\tau_d$ (with 
$\tau_d$ the deterministic time) versus $q/E$
for the same parameters of Fig.1. Inset:
the potential $U(x)$ of Eq. (2).

\noindent
FIG. 3. The shaded area is the region 
of the plane ($\ln T$, $a$) where the NES effect 
is very strong: the average escape time is greater than
$10\%$ above the deterministic escape time. 
The parameters are: $b = 7, k = 1, l = 2$.
Inset: the average escape time versus the noise intensity
for $a = 0.3$ and $T = 13.5$. The dashed
line indicates the deterministic escape time.

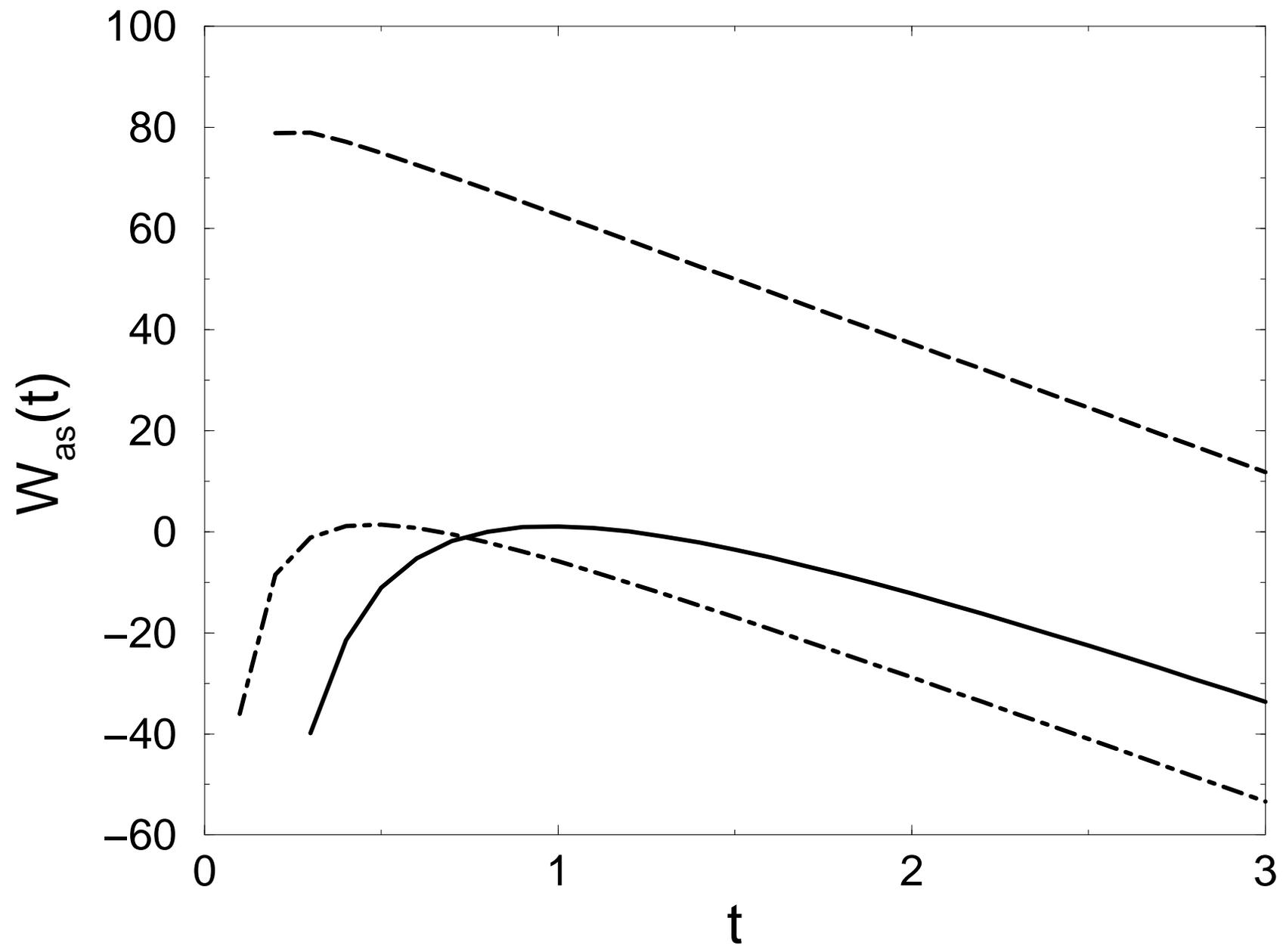
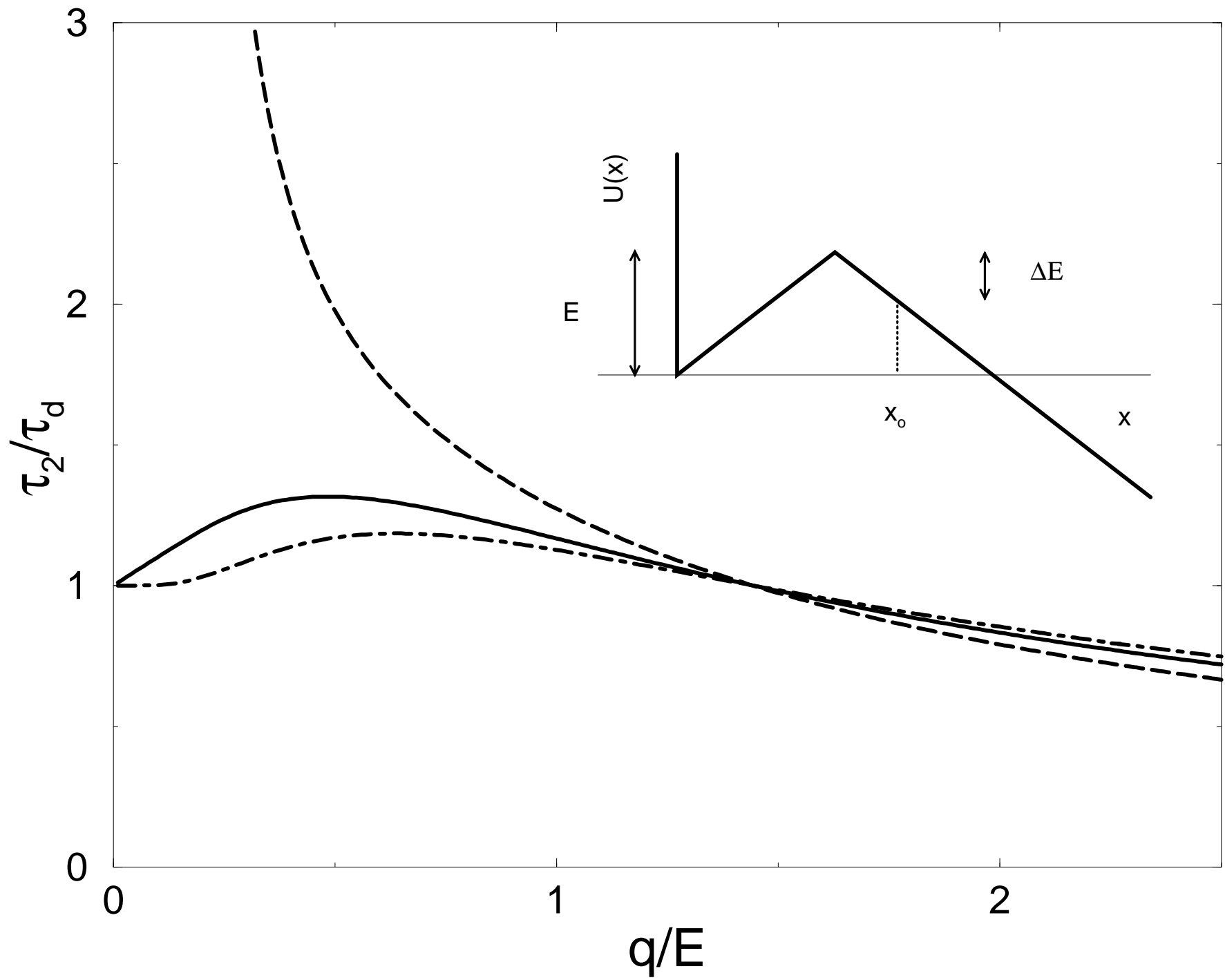
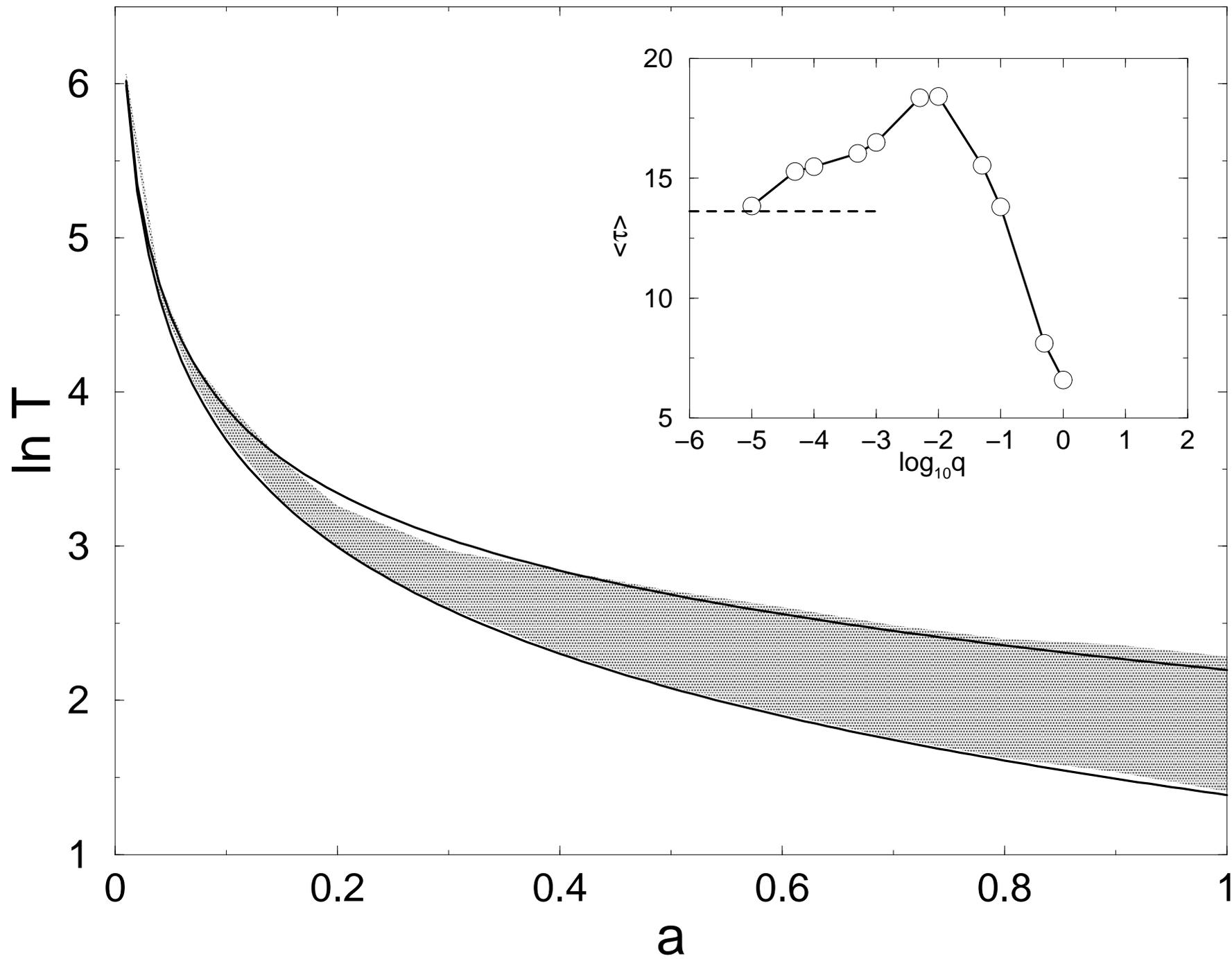
\end{document}